\documentclass[letterpaper, 10 pt, conference]{ieeeconf}  

\IEEEoverridecommandlockouts                              

\usepackage{amsmath}
\usepackage{mathtools}
\usepackage{amsfonts}
\usepackage{color,soul}
\usepackage{algorithm}
\usepackage{algpseudocode}
\usepackage{booktabs}  
\title{\LARGE \bf
Multi-Step Deep Koopman Network (MDK-Net) for Vehicle Control in Frenet Frame
}

\author{Mohammad Abtahi$^1$, Mahdis Rabbani$^1$, Armin Abdolmohammadi$^1$ and Shima Nazari$^{1}$
\thanks{*This work was not supported by any organization}
\thanks{$^{1}$ Mohammad Abtahi, Mahdis Rabbani, and Armin Abdolmohammadi are with the Mechanical and Aerospace Engineering of University of California, Davis, CA 95616, USA
        {\tt\small \{sabtahi, mrabbani, abdolmohammadi, snazari\}@ucdavis.edu}}
}

\begin{document}

\maketitle
\thispagestyle{plain}
\pagestyle{plain}

\begin{abstract}
The highly nonlinear dynamics of vehicles present a major challenge for the practical implementation of optimal and Model Predictive Control (MPC) approaches in path planning and following. Koopman operator theory offers a global linear representation of nonlinear dynamical systems, making it a promising framework for optimization-based vehicle control. This paper introduces a novel deep learning-based Koopman modeling approach that employs deep neural networks to capture the full vehicle dynamics—from pedal and steering inputs to chassis states—within a curvilinear Frenet frame. The superior accuracy of the Koopman model compared to identified linear models is shown for a double lane change maneuver. Furthermore, it is shown that an MPC controller deploying the Koopman model provides significantly improved performance while maintaining computational efficiency comparable to a linear MPC.

\end{abstract}

\section{INTRODUCTION}
Autonomous vehicles (AVs) represent a transformative technology, aiming to redefine modern transportation by enhancing safety, efficiency, and accessibility \cite{othman2022exploring}. Using advanced control strategies that govern their motion, decision making and interaction with dynamic environments, various control approaches have been developed to meet these demands, ranging from classical methods such as PID \cite{marino2011nested,kebbati2022coordinated} and LQR \cite{liu2018hierarchical} to more advanced techniques such as Model Predictive Control (MPC) \cite{abtahi2023automatic,leman2019model} and Reinforcement Learning (RL) \cite{cui2023integrated,cui2021combined}. However, designing such controllers is inherently challenging, as they need to be real-time efficient, robust to uncertainties, and highly accurate. At the core of addressing these challenges lies the importance of modeling the vehicle dynamics. These dynamics are highly nonlinear, governed by complex interactions among lateral and longitudinal motion, tire forces, and powertrain dynamics. This has motivated extensive research into nonlinear control techniques. Approaches such as Nonlinear Model Predictive Control (NMPC) \cite{jannah2022nonlinear}, Iterative Linear Quadratic Regulator (iLQR) \cite{xing2024control}, and nonlinear cascade control \cite{attia2014nonlinear} have been developed to directly address these nonlinearities. While these methods can achieve high accuracy, their computational complexity often makes them impractical for real-time applications.

Emerging AI-based methods \cite{lelko2024reinforcement} offer an alternative by learning system dynamics directly from data. These approaches have gained increasing attention for their ability to model complex nonlinear behaviors without requiring explicit system identification. However, designing such controllers remains inherently challenging due to the need for real-time efficiency, robustness to uncertainties, and high accuracy in diverse operating conditions \cite{garcia2015comprehensive}.

These limitations and challenges motivate the need for efficient and interpretable modeling frameworks that balance accuracy and computational feasibility. At the core of addressing these challenges lies the importance of linearizing the vehicle dynamic model, which serves as a foundational framework for state estimation, motion planning, and control synthesis.

One widely used approach is traditional model linearization using Taylor series expansion. This approach provides computational efficiency and fast processing speed, making it appealing for real-time control applications. However, it produces linear models that are valid only locally and fails to capture the intricate vehicle dynamics for a wide range of operating points. 

An alternative approach is the Koopman Operator, which offers a globally linear representation of a nonlinear system dynamics \cite{koopman1931hamiltonian}. In simple words, by lifting the original states of the system to a higher dimensional space using observable functions, Koopman operator captures the nonlinearities through a linear mapping on the observables \cite{brunton2021modern}. 
\begin{figure}
   \centering
   \includegraphics[width=0.8\linewidth]{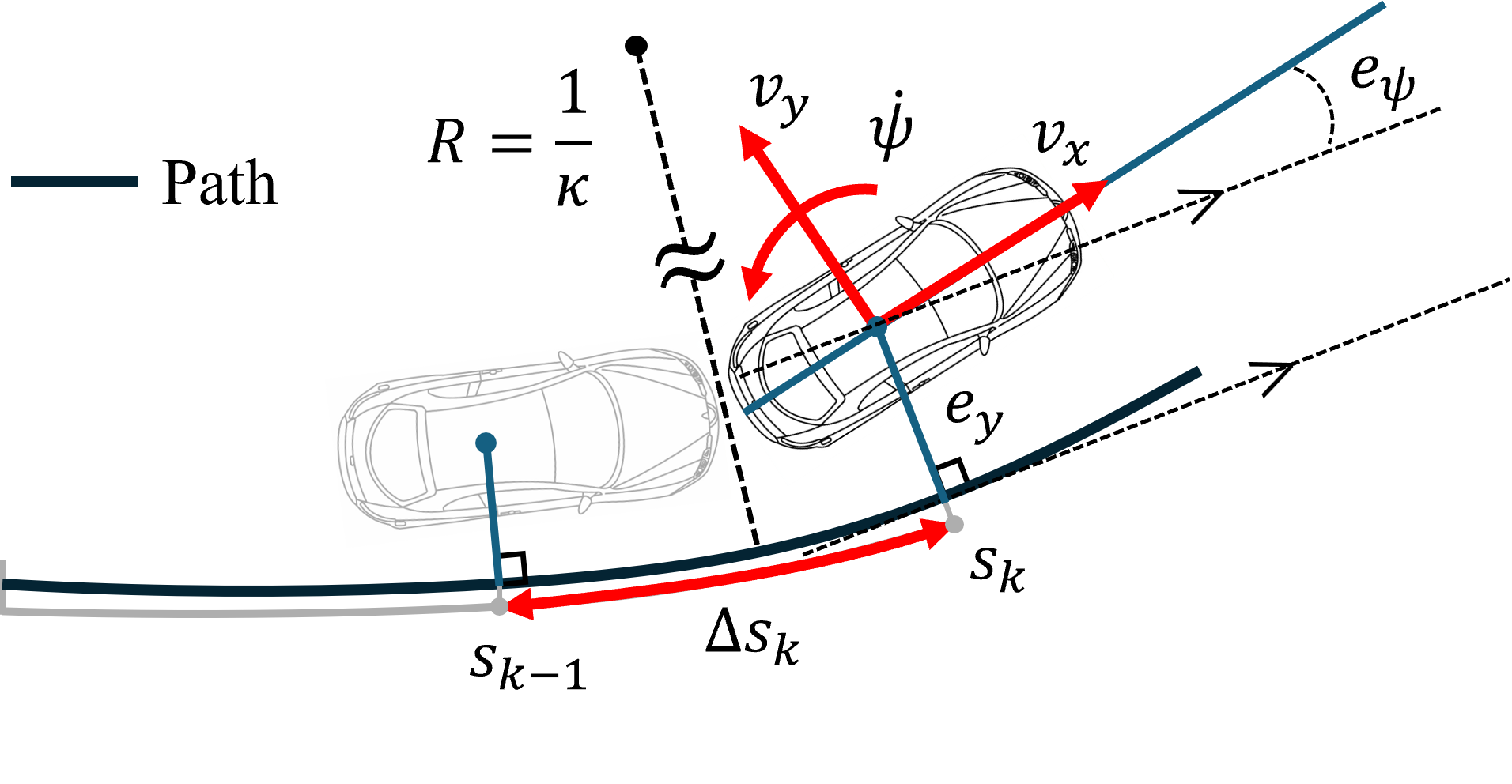} 
   \caption{Vehicle states representation in Frenet curvilinear frame.}
   \label{fig:fernet frame}
\end{figure}
The Koopman operator has received enhanced attention over the past few years, mostly due to the advances in data-driven methods and has been applied across various engineering domains, including vehicle control \cite{manzoor2023vehicular}.
It has been utilized in several applications, such as trajectory tracking using Koopman-based MPC \cite{kim2025k}, lane-keeping using Koopman-based LQR \cite{kim2022data}, and Koopman model identification in Adaptive Cruise Control (ACC) \cite{meng2024koopman}. 

Despite the promising landscape, identifying a Koopman operator for a dynamical system is challenging. The existence of a Koopman operator is only guaranteed in infinite dimension and for autonomous systems; therefore, a finite dimensional approximation for systems with control inputs might not be achievable \cite{mezic2005spectral}. 

Various data-driven methods have been proposed to produce an accurate approximation. Among them, Extended Dynamic Mode Decomposition (EDMD) with control \cite{williams2016extending} is the most widely used approach. EDMD relies on defining a rich library of manually-selected observable functions
, which act on states. The Koopman operator is then determined by solving a Mean Square Error (MSE) optimization problem \cite{brunton2021modern}. Nonetheless, selecting these observables can be challenging in practice; there are infinitely many possible choices, and determining the subset that most accurately captures the system dynamics often requires extensive trial and error. 

This challenge is universal to all EDMD-based methods and motivates using deep learning to automatically learn the set of observables rather than using a predefined set of functions. Deep learning techniques provide an end-to-end predictive model, constructed directly from data; and therefore can provide improved performance and increase the efficiency of identifying a Koopman invariant subspace \cite{shi2022deep,wang2021deep}. This motivates us to adopt a deep learning technique for our Koopman-based modeling.

Recent work by Lusch et al. \cite{lusch2018deep} has demonstrated that deep learning frameworks can be effectively employed to discover Koopman eigenfunctions. In our work, with the focus on vehicle dynamic modeling for control, we build upon several core components of the deep Koopman embedding approach. We further extend this framework by incorporating additional loss terms that enforce model stability through manipulation of the Koopman operator. We also adopt a similar approach as in \cite{shi2022deep}, jointly learning Koopman observables and the operator. In contrast, their method employs an auxiliary control network to map control inputs from the lifted space back into the original state space, whereas our approach eliminates the need for this auxiliary decoder.

On a different note, most deep Koopman-based MPC approaches for vehicle control (see \cite{xiao2023ddk,xiao2022deep}) model vehicle dynamics in a Cartesian coordinate system. In our work, we adopt a Frenet-frame coordinate system, which introduces road curvature as an exogenous input into Koopman identification. To the best of our knowledge, our work is the first application of Koopman-based modeling in vehicle dynamics that explicitly incorporates an exogenous input.

In summary, our main contributions are as follows:
\begin{itemize}
    \item To the best of our knowledge, this is the first work that applies the deep learning-based Koopman operator to vehicle dynamics in Frenet-frame coordinates, explicitly incorporating road curvature as an exogenous input,
    \item Our model utilizes low-level control inputs, i.e. steering, brake pressure and throttle, resulting in capturing powertrain and driveline dynamics, whereas previous studies have typically relied on simplified representations of powertrain and tire dynamics,
    \item We developed a novel end-to-end learning architecture that jointly learns observable functions and the Koopman operator without the need of an auxiliary decoder,
    \item Additionally, we introduce a stability loss term that penalizes any eigenvalue of the learned Koopman operator exceeding unity, thus ensuring bounded multi-step predictions for vehicle dynamics.
\end{itemize}

Fig.~\ref{fig:main_structure} illustrates the main pipeline of our work, which is organized into three key blocks. First, the \textit{Data Acquisition block} depicts how data are collected from the high-fidelity CarSim simulation environment. Second, the \textit{Multi-Step Koopman Network block} outlines our network architecture and the corresponding training process, where the Koopman operator matrices $\begin{bmatrix}A & B\end{bmatrix}$ and the encoder $\phi\left(\cdot\right)$ are trained via the acquired data simultaneously. Finally, the \textit{Simulation, Test, and Results block} demonstrates the design of an MPC controller that deploys the trained network parameters as its dynamic model for trajectory tracking in the nonlinear CarSim plant, thereby validating the effectiveness of our algorithm.

The remainder of the paper is organized to expand on these concepts in detail. Section~\ref{sec: MDK-Net} introduces the network structure, while Section~\ref{sec: training} describes the training methodology and provides a comprehensive explanation of the data collection process. Section~\ref{sec: simulation} presents the simulation setup, testing procedures, and results analysis. Finally, Section~\ref{sec: conclusion} summarizes our key findings and discusses directions for future research.

\section{Multi-step Deep Koopman Network (MDK-Net) Structure}
\label{sec: MDK-Net}
\begin{figure*}
    \centering
    \includegraphics[width=1\linewidth]{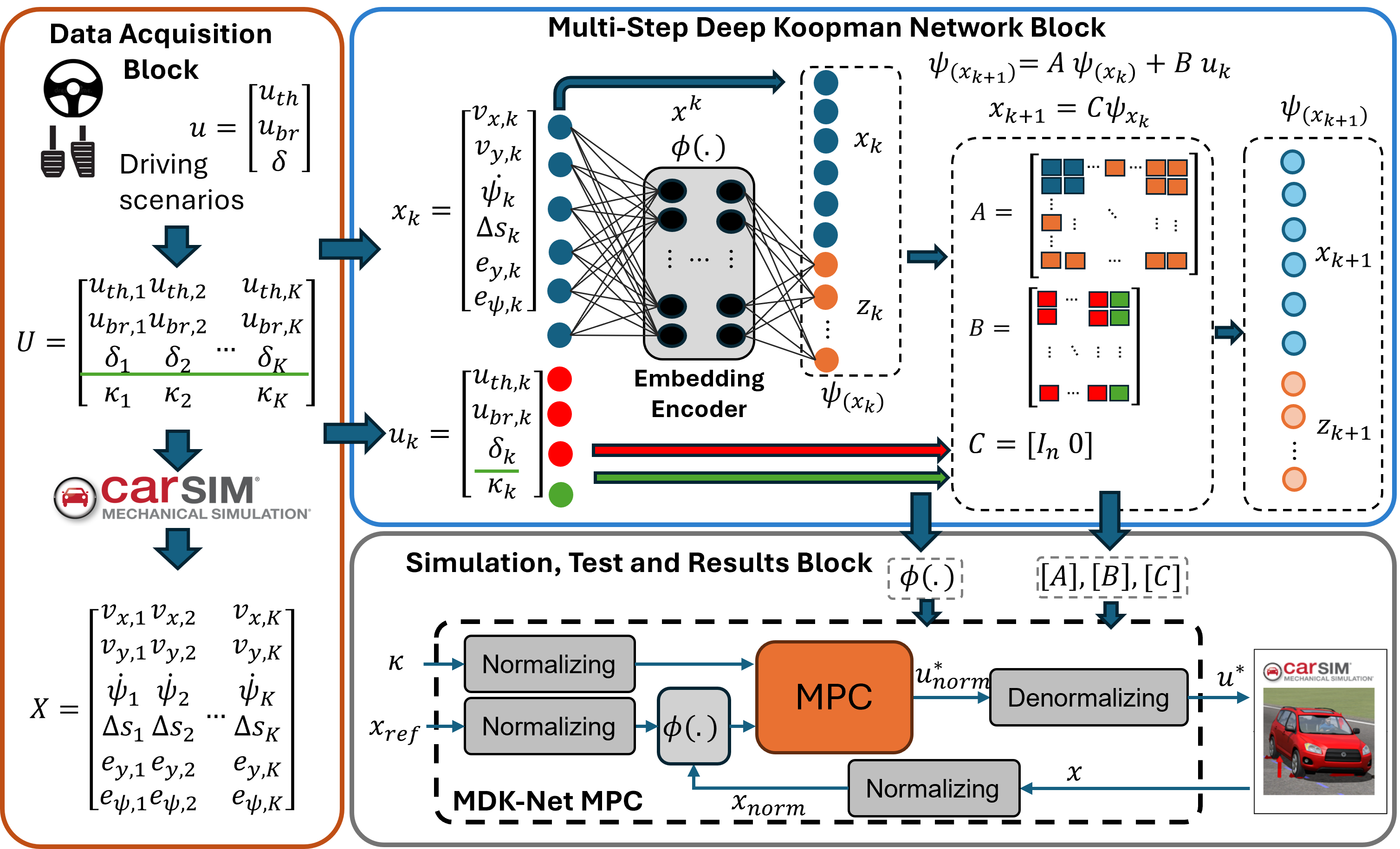}
    \caption{Overall pipeline of the proposed deep Koopman-based control framework. Driver inputs (top left) and data from CarSim (bottom left) are combined and passed through a deep neural network (center), which learns the latent states of the vehicle. The resulting Koopman matrices $A$ and $B$ capture the linear dynamics in the observable space. These matrices, together with the learned encoder $\phi(\cdot)$ and normalization layers, feed into MPC module (bottom center). The MPC generates optimal control commands, which are then fed back into the CarSim environment.}
    \label{fig:main_structure}
\end{figure*}

The general nonlinear vehicle dynamic model can be expressed as:
\begin{equation}
    x_{k+1} = \boldsymbol{F}(x_{k},\, u_{k}\bigr),
    \label{eq:vehicle_dynamics}
\end{equation}
where $x \in \mathbb{R}^n$ is the state vector with $n$ being the number of states, $u \in\mathbb{R}^{m},$ with $m$ being the number of inputs, $\boldsymbol{F}(\cdot, \cdot): \mathbb{R}^n \times \mathbb{R}^{m} \to \mathbb{R}^n$ describes the nonlinear vehicle state transition, and $k$ denotes the time step. The set of states and inputs used in this work is as follows:
\begin{gather}
\label{eq: set of states and inputs}
{x} 
= \begin{bmatrix}
v_x & v_y & \dot{\psi} & \Delta s & e_y & e_\psi
\end{bmatrix}^\top, \\
{u}
= \begin{bmatrix}
{u}_c^\top \mid {u}_w
\end{bmatrix}^\top
= \begin{bmatrix}
u_{\mathrm{th}} \quad u_{\mathrm{br}} \quad \delta \mid \kappa
\end{bmatrix}^\top.
\end{gather}

These states are defined in the Frenet curvilinear frame (as shown in Fig.~\ref{fig:fernet frame}) and correspond to the following variables: the longitudinal velocity $v_x$, lateral velocity $v_y$, yaw rate $\dot{\psi}$, progress differential $\Delta s$, lateral deviation from the reference path $e_y$, and heading error $e_\psi$. The overall control input vector is composed of $u_c$ and $u_\omega$, where the control input vector $u_c$ includes steering $\delta$, throttle $u_{th}$, and braking inputs $u_{br}$.  The variable $u_\omega = \kappa $ represents the road curvature, which is an exogenous input, measured from the environment.

For the discrete-time vehicle dynamics in (\ref{eq:vehicle_dynamics}), the Koopman operator is defined by:
\begin{equation}
    \psi(x_{k+1}) = \mathcal{K}\psi(x_k, u_k),
\end{equation}

\begin{equation}
\mathcal{K} \;=\;
\begin{bmatrix}
A & B \\[4pt]
A_{ux} & B_{uu}
\end{bmatrix},
\end{equation}
where $\mathcal{K}$ is the Koopman operator acting on a Hilbert space~$\mathcal{H}$ of observables. The sub-matrix $ \begin{bmatrix}A_{ux} & B_{uu} \end{bmatrix}$ models the evolution of the control inputs, which are not of primary interest in our identification process. Additionally, $\psi\left(x\right)$ denotes the set of observable functions. In order to obtain a finite-dimensional approximation of the Koopman operator, we approximate the dynamics by
\begin{equation}
    \psi(x_{k+1}) = A\psi(x_k) + Bu_k.
\end{equation}

In traditional approaches, the observable function $\psi(x)$ is typically defined using fixed lifting functions such as radial basis functions or polynomials in $x$. In contrast, our study learns this function directly from data using a deep neural network approach. Specifically, to construct the observables, we employ an encoder~$\phi\left(x\right)$, parameterized by $\theta$, and implemented as a fully-connected Multilayer Perceptron (MLP). This encoder maps the original state $x_k$ to a latent state space $Z_k \in \mathbb{R}^P$, where $Z_k=\phi_\theta(x_k)$. Rather than solely relying on the encoder output, we augment the observable vector by concatenating the original state with the latent states: 
\begin{equation} 
\label{eq:concatinate-latent} 
\psi(x_k) = 
\begin{bmatrix} x_k^\top \quad Z_k^\top 
\end{bmatrix}^\top. 
\end{equation}

This structure ensures that the original states can be recovered directly via a simple linear mapping:
\begin{equation}
\hat{x}_k  =  C  \psi(x_k),
\end{equation}
where $C\in \mathbb{R}^{n \times (P+n)}$ is $[I_n \ 0]$.

Our goal is to identify the Koopman operator by learning the encoder and matrices $A$ and $B$ simultaneously using a deep learning network. The MDK-Net block in Fig.~\ref{fig:main_structure} illustrates our proposed architecture, composed of a fully-connected feed-forward neural network with ReLU activation function. Matrices $A$ and $B$ serve as the final linear layers and are optimized during the training process. In the next section, we detail the loss functions used to minimize the network's error.

\subsection{Loss Functions}
\label{subsec_loss}
Unlike classical Koopman identification methods that typically focus only on one-step prediction, the MDK-Net framework is designed to penalize both single-step and multi-step prediction errors. In our formulation, a forgetting factor $\beta$ is used in the multi-step loss to gradually discount errors over the prediction horizon. This structure not only improves long-term prediction accuracy but also enforces stability on the identified dynamics by penalizing eigenvalues of the Koopman matrix \(\mathbf{A}\) that exceed the unit circle (i.e., the stability loss). In addition, a regularization loss is imposed to mitigate over-fitting of the encoder parameters. For a trajectory of length $K$, we define the following loss functions evaluated over all trajectories in a batch:
\begin{itemize}
    \item 
The single-step prediction loss $\mathcal{L}_\text{ssl}$ in the lifted space is given by:
\begin{equation}
\mathcal{L}_{\text{ssl}} = \frac{1}{K} \sum_{i=0}^{K-1} \left\| \psi({x}_{i+1}) - \left( {A}\,\psi({x}_i) + {B}\,{u}_i \right) \right\|_2^2,
\label{eq:ssl loss}
\end{equation}
where $\|\cdot\|_2$ denotes $L_2$-norm. This loss penalizes the deviation of the next step's lifted states from the one-step Koopman prediction.
\item The multi-step prediction loss 
 $\mathcal{L}_{\text{msl}}$ is formulated as:
\begin{equation}
\begin{split}
\mathcal{L}_{\text{msl}} 
&= \frac{1}{\sum_{i=1}^{K-1} \beta^i}
\sum_{i=1}^{K-1} \beta^i 
\Bigl\|\psi(x_i) \\
&\quad - \Bigl(A^{i}\,\psi(x_0) \;+\; \sum_{t=1}^{i} A^{t-1}B\,u_{i-t}\Bigr)\Bigr\|_2^2,
\end{split}
\label{eq:msl loss}
\end{equation}

Here, the error over multiple time steps is discounted by $\beta$ to prioritize near-future predictions. \item Stability loss is defined to enforce stability by penalizing any eigenvalue of $A$ with magnitude greater than one:
\begin{equation}
\mathcal{L}_{\text{sl}} = \sum_{\lambda \in \mathrm{eig}({A})} \max\bigl(0,\,|\lambda| - 1\bigr).
\label{eq:sl loss}
\end{equation}
\item  Regularization loss is applied to both the encoder parameters and the Koopman matrices to prevent overfitting:
\begin{equation}
\mathcal{L}_{\text{regl}} = \lambda_\theta \|\theta\|^2 
\;+\; \lambda_{\mathcal{K}}\left(\|{A}\|_\mathbf{F}^2 + \|{B}\|_\mathbf{F}^2\right),
\label{eq:reg loss}
\end{equation}
where $\lambda_\theta,\lambda_{\mathcal{K}}$ are the regularization coefficients and \mbox{$\|\cdot\|_{\mathbf{F}}$} denotes Frobenius norm.
\end{itemize}

The overall train loss is then defined as:
\begin{equation}
    \mathcal{L}_{\text{total}} = \alpha_1\,\mathcal{L}_{\text{ssl}} + \alpha_2\,\mathcal{L}_{\text{msl}} + \alpha_3\,\mathcal{L}_{\text{sl}} + \alpha_4\,\mathcal{L}_{\text{regl}},
    \label{eq:total loss}
\end{equation}
where the loss function weights $\alpha_i$ balance the contributions of each loss term. 

\section{Training Procedure}
\label{sec: training}

\begin{algorithm}
\caption{MDK-Net Algorithm}
\label{alg:MDK-Net}
\begin{algorithmic}[1]
\State Initialize $Epoch_{\max}$, batch size, data split, patience, $lr$, $\beta$, $\{\alpha_i\}_{i=1}^4$, $\theta$, $A$, $B$, test loss $\mathcal{L}^{\text{test}}$, best model $\Gamma$, $i\gets 0$.
\State Randomly partition the data into train and test sets and normalize them.
\While {$Epoch < Epoch_{\max}$}
    \State Randomly select a batch.
    \State Encode the state vector and concatenate it with original states as in (\ref{eq:concatinate-latent}).
    \State Compute total loss as in (\ref{eq:total loss}) and its gradient.
    \State Update $\theta$, $A$, $B$ using the computed gradient.
    \If {$Epoch$ $\mathbf{mod}\,500 == 0$}
        \State Increment $i$ by one.
        \State Evaluate total loss (\ref{eq:total loss}) on the entire test set and append it to $\mathcal{L}^{\text{test}}$.
        \If {$\mathcal{L}^{\text{test}}_i > \mathcal{L}^{\text{test}}_{i-1}>\dots>\mathcal{L}^{\text{test}}_{i-{\text{patience}}}$}
            \State Update the learning rate: $lr \gets lr \times 0.5$.
        \ElsIf {$\mathcal{L}^{\text{test}}_{i} < \mathcal{L}^{\text{test}}_{i-1}$}
            \State Update best model: $\Gamma \gets\{\theta, A, B\}$.
        \EndIf
    \EndIf
    \State Increment $Epoch$ by one.
\EndWhile
\State \textbf{Output:} Best model parameters $\Gamma=\{\theta, A, B\}$ 
\end{algorithmic}
\end{algorithm}

The training process begins with initializing key parameters such as the maximum number of epochs, batch size, learning rate ($lr$), patience, and loss function weights. 
Next, the dataset is split into train and test sets and normalized as part of the preprocessing step. Once the data are prepared, the network parameters ($\theta$, $A$ and $B$) are initialized using a Gaussian distribution. During each training iteration, a random batch of 128 trajectories is selected. 
The encoder maps the state vectors into the latent states, and these latent states are concatenated with the original states to form the observables.
The total train loss is computed on the batch. The gradient of the loss is then back-propagated to update $\theta$, $A$, and $B$. Periodically, every 500 steps, the model is evaluated on the test set. If the test loss improves over a specified patience window, the current parameters are saved as the best model; otherwise, the learning rate is decreased to promote convergence. This process continues until the maximum number of epochs is reached.
The complete MDK-Net algorithm is detailed in Algorithm~\ref{alg:MDK-Net}.

Table~\ref{tab:hyperparams} summarizes the MDK-Net training hyperparameters used in our algorithm.
\begin{table}[b]
\centering
\caption{MDK-Net Architecture \& Training Hyperparameters}
\label{tab:hyperparams}
\begin{tabular}{ll}
\hline
\textbf{Hyperparameter} & \textbf{Value} \\
\hline
Max Number of Epochs & 60,000 \\
Data Split(Train/Test) & 90\%/10\%\\
Batch Size & 128 \\
Latent Space Dimension (\(P\)) & 60 \\
Hidden Layer Sizes & [32, 64, 128, 128, 64] \\
Learning Rate (\(lr\)) & \(10^{-3}\) (reduced on plateau) \\
Patience& 3\\
Forgetting  Factor (\(\gamma\)) & 0.9 \\
Loss Function Weights (\(\alpha_1,\ldots,\alpha_4\)) & [1.0, 0.5, 1.6, $10^{-4}$] \\
$L_2$-Regularization (\(\lambda_\theta,\lambda_{\mathcal{K}}\)) & 0.9, 0.5 \\
State Dimension (\(n\)) & 6 \\
Input Dimension (\(m\)) & 4 \\
Time Step (\(\Delta t\)) & 0.025\,s \\
\hline
\end{tabular}
\end{table}

 \subsection{Data Acquisition}
\label{subsec_Data_acquisition}
As shown in the \emph{Data Acquisition} block of Fig.~\ref{fig:main_structure}, we generated our dataset using CarSim---a high-fidelity vehicle dynamics simulator that accurately replicates real-world behavior. In our experiments, we employed a C-Class hatchback car with a 150~kW engine and an all-wheel drive configuration. We simulated 7,000 episodes, each lasting 10 seconds (with a data resolution of $t_s = 25$~ms, corresponding to 400 steps per episode). To enrich the dataset and enforce diverse initial conditions, each episode was divided into five non-overlapping trajectories of 2 seconds each, yielding a total of 35,000 trajectories.

Each trajectory is characterized by a set of inputs and corresponding states. Rather than selecting inputs completely at random, we defined a library of meaningful input parameters that yield realistic driving scenarios. For example, the curvature $\kappa$ is chosen from a library of five distinct values ranging from $\kappa = -4 \times 10^{-3}$ (corresponding to a right turn with 250~m radius) to $\kappa = 4 \times 10^{-3}$ (corresponding to a left turn with a 250~m radius). The control input vector $u_c$ is constructed such that the throttle $u_{th}$ is determined within $[0,1]$, the brake $u_{br}$ is determined by a pedal pressure within $[0,150]$~N, and the steering angle $\delta$ is limited to $[-40^\circ, 40^\circ]$. Moreover, the rate of change of these inputs is constrained to match realistic driving scenarios, and only one of throttle or braking is allowed to be active at any given time.

Each test trajectory spans a full horizon of 80 steps. The procedure begins by assigning a curvature from the library to the entire horizon. Next, a time interval $\Delta T$ is selected from the dedicated library and the horizon is segmented accordingly. Within each segment, appropriate throttle and brake values are chosen from their respective libraries to ensure smooth transitions. Steering is handled separately by first selecting a specified number of steering points over the horizon and then fitting a polynomial to these points to obtain a continuous steering profile. Finally, the generated input trajectories are applied to the CarSim model.

\section{Simulation and Results}
\label{sec: simulation}
We evaluate the performance of the MDK-Net model against a Linear Time-Invariant (LTI) model obtained via system identification. Due to the inherent nonlinearities in vehicle dynamics—stemming from the complex behavior of the steering rack, powertrain, and braking systems—deriving explicit dynamic equations for a linear model is not feasible. Therefore, we use the same train and test datasets to identify a linear state-space model with an equivalent number of states as the original system via MATLAB’s System Identification Toolbox. In this process, curvature is treated as an exogenous input, influencing the identification of the $B$ matrix; consequently, multiple LTI models are identified for different curvature ranges, as described in Section~\ref{subsec_Data_acquisition}. The performance of the MDK-Net and LTI models is then compared for each curvature value by computing the average MSE over 2,400 open-loop test trajectories.

Table~\ref{tab:mse_comparison} presents the results over an 80-step (2-second) horizon for all the aforementioned test trajectories, indicating that the MDK-Net achieves significantly lower MSE across all state variables compared to the LTI models. In particular, for critical lateral dynamics such as \(v_y\) and \(e_y\), the MDK-Net’s multi-step prediction capability enables a notably more accurate performance. 
\begin{table}[b]
    \centering
    \setlength{\tabcolsep}{4pt}  
    \caption{Average open loop 80 step (2 second) prediction MSE for 2400 test trajectories}
    \label{tab:mse_comparison}
    \begin{tabular}{lcccccc}
        \toprule
        \textbf{Model} & $V_x$ & $V_y$ & $\dot{\psi}$ & $\Delta s$ & $e_y$ & $e_\psi$ \\
        & (km/h) & (km/h) & (deg/s) & (m) & (m) & (deg) \\
        \midrule
        \textbf{MDK-Net}  & \textbf{2.908}  & \textbf{0.020}  & \textbf{0.445}  & \textbf{0.00045}  & \textbf{2.504}  & \textbf{0.761}  \\
        \textbf{LTI model} & 6937.684  & 0.548  & 3.139  & 0.2305  & 39.303  & 11.242  \\
        \bottomrule
    \end{tabular}
\end{table}

To further validate the predictive accuracy of our \mbox{ MDK-Net} model, we compare its open-loop performance against the full nonlinear vehicle model simulated in CarSim in a double lane change scenario inside a constant curvature of $\kappa= 0.001$. Identical input trajectories are applied to both the MDK-Net model and the LTI model, with the outputs then compared to those of the nonlinear CarSim model. 

\begin{figure}[t]
   \centering
   \includegraphics[width=1\linewidth]{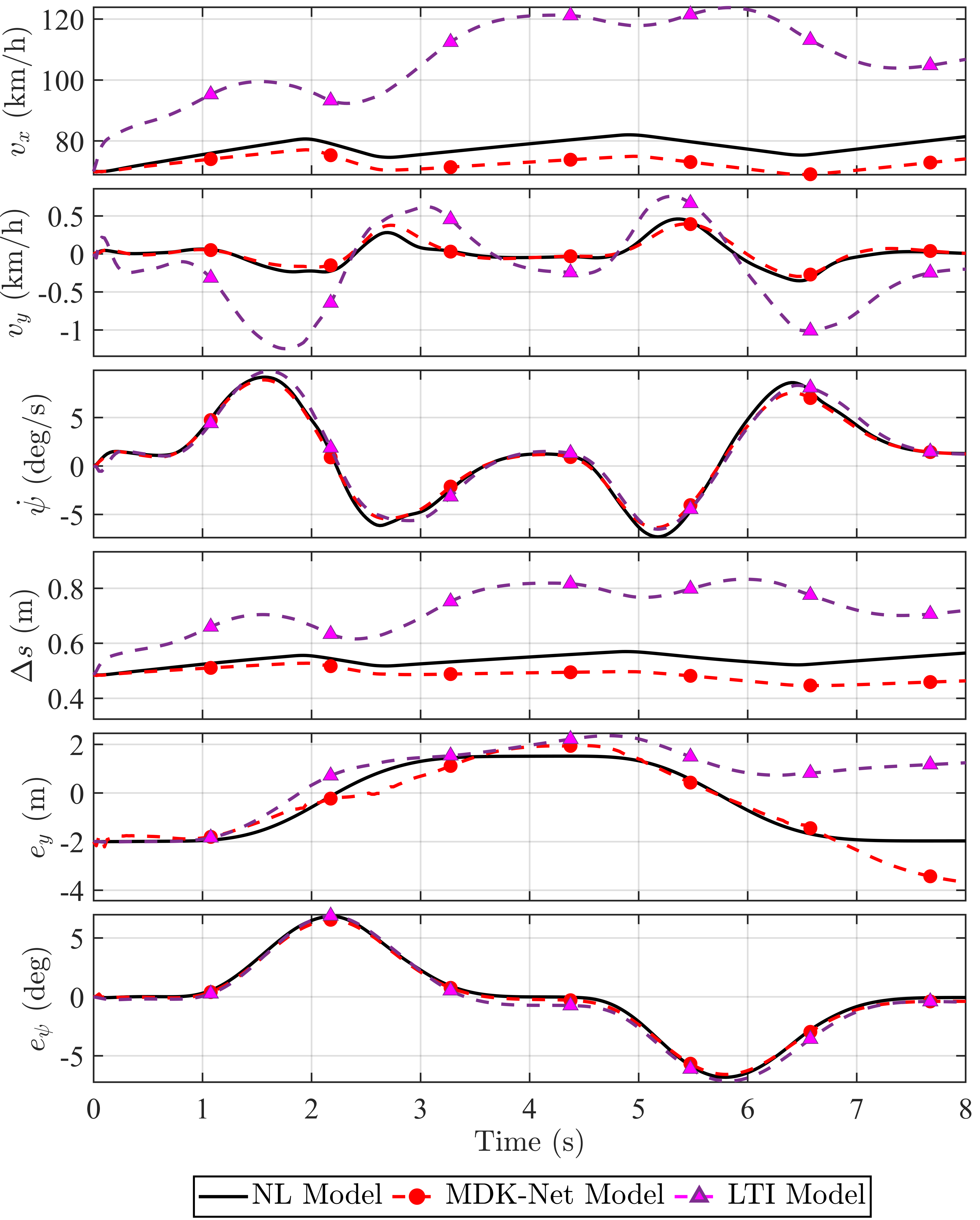} 
   \caption{Comparison of open-loop state trajectories between the Reference (Nonlinear model), MDK-Net (Deep Koopman model), and LTI model (MATLAB System Identification-based linear model) for a double lane change scenario.}
   \label{fig:trajectory_comparison}
\end{figure}

As illustrated in Fig.~\ref{fig:trajectory_comparison}, the MDK-Net model accurately captures the evolution of all state variables, notably the lateral dynamics, $v_y$ and $e_y$ and longitudinal dynamics, $v_x$ and $\Delta s$, with significantly higher precision than those of the LTI model.
\subsection{Trajectory Tracking via MPC}
\label{subsec_MPC}
MPC has received significant recognition for its capability to address both linear and nonlinear control problems \cite{rokonuzzaman2021model}. It operates by repeatedly solving an optimization problem in a receding-horizon manner and applying the first control action at each iteration. In the remainder of this section, we design an MPC controller that utilizes the Koopman model identified through our proposed MDK-Net for trajectory tracking to evaluate its accuracy and computational efficiency. 

Since the Koopman operator acts on the observable space, the corresponding optimization problem must also be formulated in this space. Thus, we utilize the same observables defined in (\ref{eq:concatinate-latent}) as the optimization states. Furthermore, we decompose the vector $u$ in (\ref{eq: set of states and inputs}) back into the control input $u_c$ and the exogenous input $u_w$. Consequently, the Koopman block matrix $B$, trained through the network, must also be decomposed as $B=\begin{bmatrix}B_{u_c} \mid B_{u_w}\end{bmatrix}$. During optimization, the MPC adjusts only the block corresponding to the actual control inputs, treating the curvature block as an exogenous input.

The optimal control sequence $\{u_{c,k}\}_{k=0}^{N_p}$ is determined by solving the following optimization problem:
\begin{equation}
\label{eq:mpc_optimization}
\begin{aligned}
\underset{\{u_{c,k}\}_{k=0}^{N_p}}{\operatorname{min}} \quad 
& \sum_{k=0}^{N-1} \Bigl[ (\psi_k - \psi_r)^\top Q (\psi_k - \psi_r) + u_{c,k}^\top R u_{c,k} \Bigr] \\
& + (\psi_N - \psi_r)^\top Q_N (\psi_N - \psi_r) \\
\text{subject to} \quad 
& \psi_{k+1} = A \psi_k + B_{u_c} u_{c,k} + B_{u_w} u_{w,k},  \\
& u_{\min} \leq u_{c,k} \leq u_{\max}, \\
& \psi_0 = \psi(x_0), \quad k = 0, \dots, N-1,
\end{aligned}
\end{equation}
where $Q, Q_N$ are the state weighting matrices, $R$ denotes the input weighting matrix, and $\psi_r$ is the lifted reference states. In addition, $u_{w,k}$ is assumed to be constant over the entire horizon.

To validate the performance of our designed MPC controller, we adopted the same double lane change scenario used in the model comparison. The MDK-Net–based MPC controller was implemented using a Simulink model that interacts with a high-fidelity CarSim plant and follows a reference trajectory with a constant curvature of $\kappa= 0.001$. For comparison, we also designed an MPC controller based on an identified LTI model using MATLAB’s MPC Toolbox. 

\begin{figure}[t]
   \centering
   \includegraphics[width=1\linewidth]{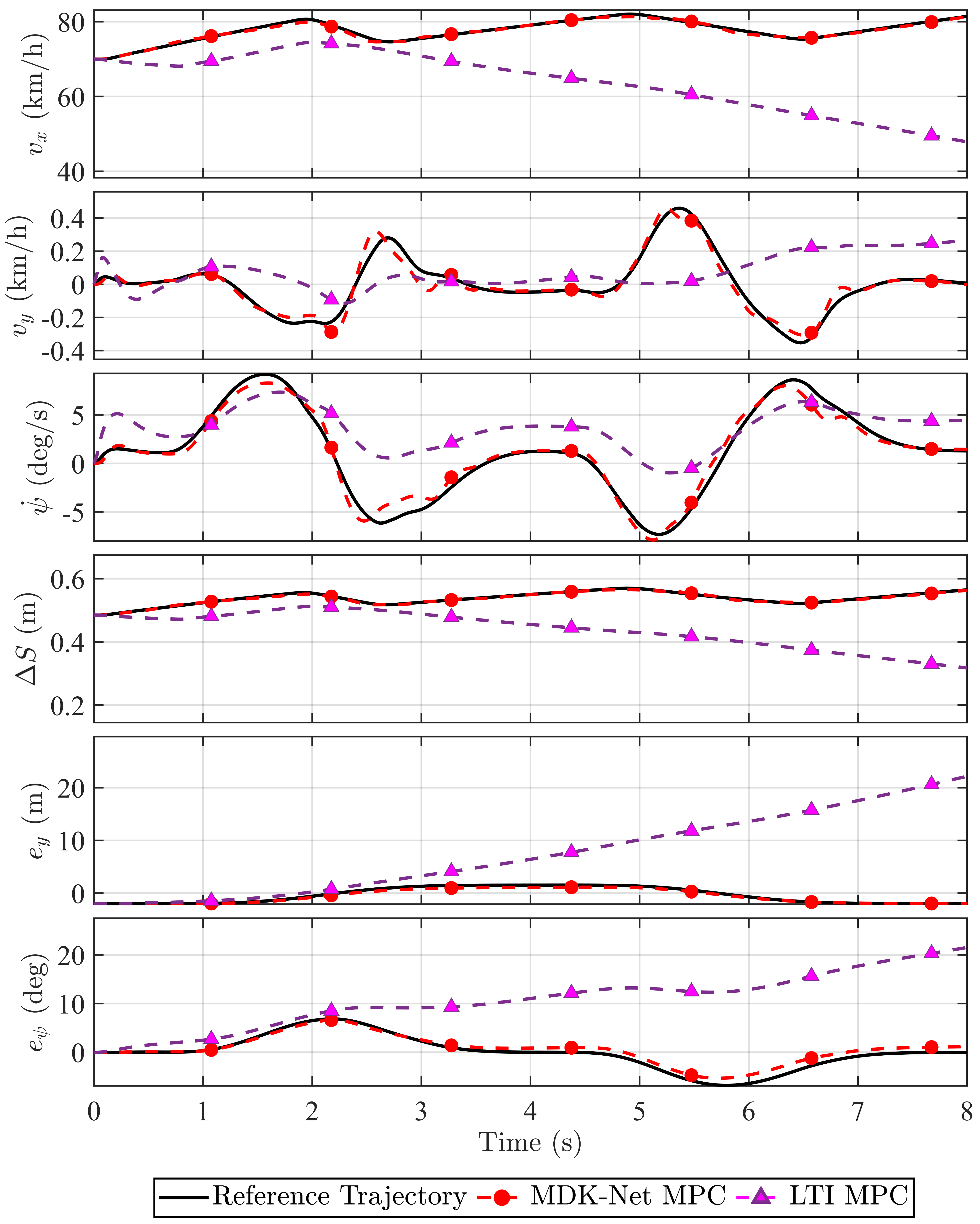} 
   \caption{Trajectory tracking performance of MDK-Net compared to LTI MPC controller which uses the 6 state model identified by MATLAB System Identification. MDK-net MPC was able to successfully track the reference trajectory with minimal error compared to the other controller.}
   \label{fig:MPC_comparison}
\end{figure}

Fig.~\ref{fig:MPC_comparison} compares the performance of these two controllers. As illustrated, the MDK-Net–based MPC tracks the reference trajectory with high accuracy across all state variables, whereas the LTI-based controller exhibits noticeable drift from reference—particularly in lateral and heading error after the first lane change. Furthermore, the LTI controller fails to adequately track the lateral and angular velocities, which are highly dependent on the nonlinear relationship between the steering input and the vehicle dynamics. This underscores the need for a model that compensates for these nonlinearities through a higher-dimensional linear approximation.


In terms of computational performance, the convexity of the MDK-Net–based MPC optimization problem ensures that its computation time is comparable to that of the LTI-based MPC, despite the higher dimensionality of the lifted state. For a double lane change scenario simulation of length 10 seconds, with a time step of $t_s=25$~ms and a controller prediction horizon of $N_p=20$, the total MPC computation time was 3.2 seconds for the LTI MPC and 7.4 seconds for the MDK-Net MPC, on a system equipped with an 11th Gen Intel\(^{\textregistered}\) Core\(^{\text{TM}}\) i7-1165G7 CPU @ 2.80 GHz (4 cores, 8 logical processors). 

These results indicate that the proposed approach is viable for real-time implementation, especially when contrasted with conventional nonlinear MPC schemes that typically demand significantly higher computational resources.


\section{Conclusion}
\label{sec: conclusion}
This work proposed a deep Koopman framework for modeling vehicle dynamics in the curvilinear Frenet frame. The model takes pedal and steering inputs as control actuators and incorporates road curvature as an exogenous input. Unlike prior approaches that use a library of predefined observable functions and optimize single-step predictions, the proposed framework jointly learns the lifting functions and Koopman matrices while minimizing trajectory-level prediction errors. High-fidelity vehicle simulations demonstrate that the developed Koopman model significantly outperforms linear time-invariant models trained on the same dataset in terms of prediction accuracy. Furthermore, an MPC-based path-following controller utilizing the Koopman model was tested in a double lane-change scenario, showing superior performance while maintaining computational efficiency comparable to linear MPC.

Future research directions include extending the developed model for path planning applications. Additionally, linearizing nonlinear state-dependent constraints, such as collision avoidance constraints, for automated driving and racing presents another promising avenue. Finally, exploring more advanced Koopman structures, such as bilinear models, could further enhance prediction accuracy.

\section*{ACKNOWLEDGMENT}
We would like to acknowledge the assistance of ChatGPT by OpenAI \cite{ChatGPT} in refining the manuscript’s clarity, grammar, and overall coherence. However, all results, figures, and graphs in this work were generated entirely without any AI assistance.

\bibliographystyle{IEEEtran}
\bibliography{references}

\begin{thebibliography}{10}
\providecommand{\url}[1]{#1}
\csname url@samestyle\endcsname
\providecommand{\newblock}{\relax}
\providecommand{\bibinfo}[2]{#2}
\providecommand{\BIBentrySTDinterwordspacing}{\spaceskip=0pt\relax}
\providecommand{\BIBentryALTinterwordstretchfactor}{4}
\providecommand{\BIBentryALTinterwordspacing}{\spaceskip=\fontdimen2\font plus
\BIBentryALTinterwordstretchfactor\fontdimen3\font minus \fontdimen4\font\relax}
\providecommand{\BIBforeignlanguage}[2]{{%
\expandafter\ifx\csname l@#1\endcsname\relax
\typeout{** WARNING: IEEEtran.bst: No hyphenation pattern has been}%
\typeout{** loaded for the language `#1'. Using the pattern for}%
\typeout{** the default language instead.}%
\else
\language=\csname l@#1\endcsname
\fi
#2}}
\providecommand{\BIBdecl}{\relax}
\BIBdecl

\bibitem{othman2022exploring}
K.~Othman, ``Exploring the implications of autonomous vehicles: A comprehensive review,'' \emph{Innovative Infrastructure Solutions}, vol.~7, no.~2, p. 165, 2022.

\bibitem{marino2011nested}
R.~Marino, S.~Scalzi, and M.~Netto, ``Nested pid steering control for lane keeping in autonomous vehicles,'' \emph{Control Engineering Practice}, vol.~19, no.~12, pp. 1459--1467, 2011.

\bibitem{kebbati2022coordinated}
Y.~Kebbati, N.~Ait-Oufroukh, V.~Vigneron, and D.~Ichalal, ``Coordinated pso-pid based longitudinal control with lpv-mpc based lateral control for autonomous vehicles,'' in \emph{2022 European Control Conference (ECC)}.\hskip 1em plus 0.5em minus 0.4em\relax IEEE, 2022, pp. 518--523.

\bibitem{liu2018hierarchical}
Q.~Liu, Y.~Liu, C.~Liu, B.~Chen, W.~Zhang, L.~Li, and X.~Ji, ``Hierarchical lateral control scheme for autonomous vehicle with uneven time delays induced by vision sensors,'' \emph{Sensors}, vol.~18, no.~8, p. 2544, 2018.

\bibitem{abtahi2023automatic}
M.~Abtahi, M.~Rabbani, and S.~Nazari, ``An automatic tuning mpc with application to ecological cruise control,'' \emph{IFAC-PapersOnLine}, vol.~56, no.~3, pp. 265--270, 2023.

\bibitem{leman2019model}
Z.~A. Leman, M.~H.~M. Ariff, H.~Zamzuri, M.~A.~A. Rahman, and S.~A. Mazlan, ``Model predictive controller for path tracking and obstacle avoidance manoeuvre on autonomous vehicle,'' in \emph{2019 12th Asian Control Conference (ASCC)}.\hskip 1em plus 0.5em minus 0.4em\relax IEEE, 2019, pp. 1271--1276.

\bibitem{cui2023integrated}
J.~Cui, B.~Zhao, and M.~Qu, ``An integrated lateral and longitudinal decision-making model for autonomous driving based on deep reinforcement learning,'' \emph{Journal of Advanced Transportation}, vol. 2023, no.~1, p. 1513008, 2023.

\bibitem{cui2021combined}
L.~Cui, K.~Ozbay, and Z.-P. Jiang, ``Combined longitudinal and lateral control of autonomous vehicles based on reinforcement learning,'' in \emph{2021 American control conference (ACC)}.\hskip 1em plus 0.5em minus 0.4em\relax IEEE, 2021, pp. 1929--1934.

\bibitem{jannah2022nonlinear}
S.~W. Jannah and A.~Santoso, ``Nonlinear model predictive control for longitudinal and lateral dynamic of autonomous car,'' in \emph{2022 11th Electrical Power, Electronics, Communications, Controls and Informatics Seminar (EECCIS)}.\hskip 1em plus 0.5em minus 0.4em\relax IEEE, 2022, pp. 145--148.

\bibitem{xing2024control}
J.~Xing, Z.~Song, M.~Cheng, Q.~Gou, X.~Bai, and S.~Chen, ``Control of vehicle trajectory tracking under extreme conditions based on ilqr,'' in \emph{2024 9th International Conference on Electronic Technology and Information Science (ICETIS)}.\hskip 1em plus 0.5em minus 0.4em\relax IEEE, 2024, pp. 254--258.

\bibitem{attia2014nonlinear}
R.~Attia, R.~Orjuela, and M.~Basset, ``Nonlinear cascade strategy for longitudinal control in automated vehicle guidance,'' \emph{Control Engineering Practice}, vol.~29, pp. 225--234, 2014.

\bibitem{lelko2024reinforcement}
A.~Lelk{\'o}, B.~N{\'e}meth, and P.~G{\'a}sp{\'a}r, ``Reinforcement learning-based robust vehicle control for autonomous vehicle trajectory tracking,'' \emph{Engineering Proceedings}, vol.~79, no.~1, p.~30, 2024.

\bibitem{garcia2015comprehensive}
J.~Garc{\i}a and F.~Fern{\'a}ndez, ``A comprehensive survey on safe reinforcement learning,'' \emph{Journal of Machine Learning Research}, vol.~16, no.~1, pp. 1437--1480, 2015.

\bibitem{koopman1931hamiltonian}
B.~O. Koopman, ``Hamiltonian systems and transformation in hilbert space,'' \emph{Proceedings of the National Academy of Sciences}, vol.~17, no.~5, pp. 315--318, 1931.

\bibitem{brunton2021modern}
\BIBentryALTinterwordspacing
S.~L. Brunton, M.~Budi\v{s}i\'{c}, E.~Kaiser, and J.~N. Kutz, ``Modern koopman theory for dynamical systems,'' \emph{SIAM Review}, vol.~64, no.~2, pp. 229--340, 2022. [Online]. Available: \url{https://doi.org/10.1137/21M1401243}
\BIBentrySTDinterwordspacing

\bibitem{manzoor2023vehicular}
W.~A. Manzoor, S.~Rawashdeh, and A.~Mohammadi, ``Vehicular applications of koopman operator theory—a survey,'' \emph{IEEE Access}, vol.~11, pp. 25\,917--25\,931, 2023.

\bibitem{kim2025k}
J.~S. Kim, Y.~S. Quan, C.~C. Chung, and W.~Y. Choi, ``K-smpc: Koopman operator-based stochastic model predictive control for enhanced lateral control of autonomous vehicles,'' \emph{IEEE Access}, 2025.

\bibitem{kim2022data}
J.~S. Kim, Y.~S. Quan, and C.~C. Chung, ``Data-driven modeling and control for lane keeping system of automated driving vehicles: Koopman operator approach,'' in \emph{2022 22nd international conference on control, automation and systems (ICCAS)}.\hskip 1em plus 0.5em minus 0.4em\relax IEEE, 2022, pp. 1049--1055.

\bibitem{meng2024koopman}
Y.~Meng, H.~Li, M.~Ornik, and X.~Li, ``Koopman-based data-driven techniques for adaptive cruise control system identification,'' in \emph{27th IEEE International Conference on Intelligent Transportation Systems (ITSC), IEEE}, 2024.

\bibitem{mezic2005spectral}
I.~Mezi{\'c}, ``Spectral properties of dynamical systems, model reduction and decompositions,'' \emph{Nonlinear Dynamics}, vol.~41, pp. 309--325, 2005.

\bibitem{williams2016extending}
M.~O. Williams, M.~S. Hemati, S.~T. Dawson, I.~G. Kevrekidis, and C.~W. Rowley, ``Extending data-driven koopman analysis to actuated systems,'' \emph{IFAC-PapersOnLine}, vol.~49, no.~18, pp. 704--709, 2016.

\bibitem{shi2022deep}
H.~Shi and M.~Q.-H. Meng, ``Deep koopman operator with control for nonlinear systems,'' \emph{IEEE Robotics and Automation Letters}, vol.~7, no.~3, pp. 7700--7707, 2022.

\bibitem{wang2021deep}
R.~Wang, Y.~Han, and U.~Vaidya, ``Deep koopman data-driven optimal control framework for autonomous racing,'' \emph{Early Access}, vol.~5, pp. 1--8, 2021.

\bibitem{lusch2018deep}
B.~Lusch, J.~N. Kutz, and S.~L. Brunton, ``Deep learning for universal linear embeddings of nonlinear dynamics,'' \emph{Nature communications}, vol.~9, no.~1, p. 4950, 2018.

\bibitem{xiao2023ddk}
Y.~Xiao, X.~Zhang, X.~Xu, Y.~Lu, and J.~Lil, ``Ddk: A deep koopman approach for longitudinal and lateral control of autonomous ground vehicles,'' in \emph{2023 IEEE International Conference on Robotics and Automation (ICRA)}.\hskip 1em plus 0.5em minus 0.4em\relax IEEE, 2023, pp. 975--981.

\bibitem{xiao2022deep}
Y.~Xiao, X.~Zhang, X.~Xu, X.~Liu, and J.~Liu, ``Deep neural networks with koopman operators for modeling and control of autonomous vehicles,'' \emph{IEEE transactions on intelligent vehicles}, vol.~8, no.~1, pp. 135--146, 2022.

\bibitem{rokonuzzaman2021model}
M.~Rokonuzzaman, N.~Mohajer, S.~Nahavandi, and S.~Mohamed, ``Model predictive control with learned vehicle dynamics for autonomous vehicle path tracking,'' \emph{IEEE Access}, vol.~9, pp. 128\,233--128\,249, 2021.

\bibitem{ChatGPT}
\BIBentryALTinterwordspacing
OpenAI, ``Chatgpt 4.0,'' Large language model, 2025. [Online]. Available: \url{https://chat.openai.com/chat}
\BIBentrySTDinterwordspacing

\end{thebibliography}

\end{document}